\documentstyle[aps,preprint]{revtex}
\begin{document}
\draft
\preprint{}

\title{Energy spectrum for two-dimensional potentials in very high 
magnetic fields}

\author{Z. Gedik and M. Bayindir}

\address{Department of Physics, Bilkent University, 
Bilkent 06533 Ankara, Turkey}

\date{\today}
\maketitle
\begin{abstract}

A new method, analogous to supersymmetry transformation in quantum 
mechanics, is developed for a particle in the lowest Landau level moving 
in an arbitrary potential. The method is applied to two-dimensional
potentials formed by Dirac delta scattering centers. In the periodic case, 
the problem is solved exactly for rational values of the magnetic 
flux (in units of flux quantum) per unit cell. The spectrum is found to 
be self-similar, resembling the Hofstadter butterfly\cite{hofstadter}. 

\end{abstract}


\narrowtext

In recent years, the energy spectrum of two-dimensional (2D) electron
systems have attracted great interest, because of the relevance of the
problem to the magneto-transport properties\cite{daniela} and in particular 
to the quantum Hall effect\cite{chakraborty}. It is 
believed that the physics of the integer quantum Hall effect is governed 
by the interaction 
of electrons with a disordered potential which leads to localization of 
the eigenfunctions. Observed conductance steps can be explained by a
sequence of localization-delocalization transitions. Quantization of the 
Hall conductance due to periodic potentials has also been 
studied\cite{streda,thouless}. In this case, the presence of steps is 
explained by the gaps in the energy spectrum. 

There are two opposite approaches to the problem of an electron moving in a
periodic potential: The tight-binding and the nearly-free-electron methods.
In the first approach, the magnetic field is introduced via Peierls
substitution where the matrix elements are multiplied by 
$\exp[\frac{iq}{\hbar c}\int{\bf A}\cdot d{\bf l}]$\cite{peierls}. On the other
hand, in the case of nearly-free-electrons, Landau level structure is
essential and the lattice potential is introduced via intra- and
inter-Landau level scattering matrix elements. The duality between the 
position and the momentum in quantum mechanics leads to similarities between 
the two methods. In the presence of magnetic field, the secular equations 
for the two limits, the tight-binding and the nearly-free-electron approach, 
with certain approximations, are identical\cite{langbein}. 
The characteristic feature of the problem is that the secular determinant 
for the limit of infinite crystals can be reduced to a finite determinant, 
when the magnetic flux per unit cell is a rational number in units of the 
flux quantum. Tight-binding case has been studied by 
Azbel'\cite{azbel} and Hofstadter\cite{hofstadter}, who showed that 
the system has a complicated self-similar spectrum.

In this work, a new approach is developed for a particle in the lowest
Landau level moving in an arbitrary potential. The method is used to obtain 
the energy spectrum in the presence of Dirac delta potentials. 
The difficulty with Dirac delta potential is that, even if
the inter-Landau level couplings are small and can be neglected, there is a 
strong intra-Landau level mixing. The formalism developed in this study, 
leads to an eigenvalue problem where the coupling between distant sites
become negligibly small. Moreover, since the problem is formulated in real
space, the distribution of Dirac delta potentials can be arbitrary. 
In spite of the sharpness of the potential, the assumption that particle 
is confined into a single Landau level is consistent with the result that 
shifts in energy eigenvalues are very small in comparison to inter-Landau 
level separation. The method can be used for any potential, as far as the
particle is confined into the lowest Landau level. After presenting the
method, the case of Dirac delta scattering centers will be solved for 
periodic distribution. 

The Hamiltonian for a particle of mass $m$ and charge $q$, moving in 2D in
the presence of magnetic field ${\bf B}={\bf \nabla}\times{\bf A}$
perpendicular to the plane and potential $V$, is given by $H=H_{0}+V$ where 
\begin{equation}
H_{0}=\frac{1}{2m}({\bf p}-\frac{q}{c}{\bf A})^{2}~~~.
\end{equation}
Using the symmetric gauge 
${\bf A}=\frac{1}{2}{\bf B}\times{\bf r}$ and complex coordinates 
$z=X+iY=\sqrt{qB/2\hbar c}(x+iy)$ where ${\bf r}=(x,y)$, the unperturbed
Hamiltonian can be written as $H_0=a^\dagger a$ where
$a^{\dagger}=-\partial/\partial z+z^*/2$. Since 
$[a, a^\dagger ] = 1$, the energy eigenvalues are given
by $E_n=\hbar\omega(n+1/2)$ where $\omega=qB/2mc$ ($q$ is 
assumed to be positive) and $n$= 0, 1, 2, ....
The ground state wave
functions are of the form $f(z)e^{-|z|^2/2}$. Here $f(z)$ is any analytic
function of $z$. A basis for the infinitely 
degenerate ground state is formed by 
\begin{equation}
\phi_m(z,z^*)=\frac{z^m}{\sqrt{\pi m!}}e^{-|z|^2/2}
\end{equation}
with $m=0, 1, 2, ...$. 
When the magnetic field is very high the particle
is confined into the lowest Landau level. This is a good 
approximation as long as the potential is small in comparison to 
Landau level splitting $\hbar\omega$. 
The problem is reduced to the 
diagonalization of the matrix $V$ whose entries are given 
by
\begin{equation}
<\!\!m|V|m'\!\!>=\int\!\!\int \phi_m^*(z,z^*)V(z, z^*)\phi_m(z,z^*)d^2z ~~~,
\end{equation}
where $d^2z=dx dy$. Let $A$ and $B$ be defined by
\begin{eqnarray}
A_{m,zz^*}=&\phi_m^*(z,z^*)\sqrt{V(z, z^*)} \\ \nonumber
B_{zz^*,m}=&\sqrt{V(z, z^*)}\phi_m(z,z^*) ~~~.
\end{eqnarray}
Therefore $V=AB$. 
For positive $V(z, z^*)$, $A=B^\dagger$. 
Now consider the operator 
${\tilde V}=BA$. Any eigenvalue $E$ associated with the eigenket
$|{\tilde \psi}\!\!>$ of the operator ${\tilde V}=BA$ is also an
eigenvalue of the operator $V=AB$, except when 
$A|{\tilde \psi}\!\!>=0$. The corresponding ket is given by 
$|\psi\!\!>=E^{-1/2}A|{\tilde \psi}\!\!>$. Similarly, 
any eigenvalue $E$ associated with the eigenket
$|\psi\!\!>$ of the operator $V=AB$ is also an
eigenvalue of the operator ${\tilde V}=BA$, except when 
$B|\psi\!\!>=0$ and the corresponding ket is given by 
$|{\tilde \psi}\!\!>=E^{-1/2}B|\psi\!\!>$. Therefore it is enough to 
diagonalize
${\tilde V}$, instead of $V$, to find the energy eigenvalues, except for 
the ground state. This method is analogous to supersymmetric quantum
mechanics pioneered by Witten\cite{witten}. Substitution of $A$ and 
$B$ yields
\begin{equation}
<\!\!zz^*|{\tilde V}|z'z'^*\!\!>=\sqrt{\frac{V(z,z^*)}{\pi}}
e^{zz'^*-|z|^2/2-|z'|^2/2}\sqrt{\frac{V(z',z'^*)}{\pi}}~~~.
\end{equation}
When $V(z,z^*)$ is negative ${\tilde V}=BA$ is not Hermitian but the
eigenvalues are still real. 
The method can be applied to any problem in the lowest Landau level 
but its advantages will be demonstrated for the potential 
\begin{equation}
V({\bf r})=V_0\sum_{i} \delta({\bf r}-{\bf r}_i)~~~.
\end{equation}
In this case, $z$ and $z'$ become discrete variables and the matrix 
to be diagonalized is
\begin{equation}
<\!\!i|{\tilde V}|j\!\!>=\frac{V_0}{\pi}e^{z_iz_j^*-|z_i|^2/2-|z_j|^2/2}~~~.
\end{equation}
It is interesting to note that the matrix elements are very similar to those
of the tight-binding problem. Phase is nothing but the usual Peierls
factor. 
$<\!\!i|{\tilde V}|j\!\!>$ and $<\!\!m|V|m'\!\!>$ have the same spectrum 
except for the
ground state. Let $N_i$ be the number of impurities and $N$ be the 
number of states both of which go to infinity so that 
$N_i/N\rightarrow\rho$. Hence ${\tilde V}$ and $V$ have $N_i$ and $N$
eigenvalues, respectively. The bigger one has 
$|N_i-N|$ zero eigenvalues. If $N_i < N$, $N-N_i$ eigenvalues of $V$ will be
zero. ${\tilde V}$ {\em filters} the nonzero eigenvalues of $V$. 
However, if ${\tilde V}$ itself gives zero eigenvalues, the situation must
be examined carefully to distinguish the zero eigenvalues from the extra
zeros.
Once 
${\tilde V}$ is diagonalized, the eigenfunctions of $V$ can be
constructed by acting $A$ on the eigenvectors of ${\tilde V}$. 
The impurities can be distributed in an arbitrary way. 

The problem of an electron in a magnetic field interacting with point
impurities has been discussed extensively in the
literature\cite{prange,ando,avishai,huang,brako,ishikawa}. 
It has been shown that the
zeros of the wave function can be adjusted to coincide with the locations of
the scatterers if the concentration of the scatterers is low enough. 
This corresponds
to $N_i < N$ case discussed above. The presence of $N-N_i$
energy eigenvalues unaffected by the impurities is a direct consequence of
the dimensionality of the matrix to be diagonalized. 
Furthermore, the method gives the full energy spectrum for any distribution 
of impurities. It must be noted that
Dirac delta potentials have some peculiarities in 2D. They 
do not cause scattering, irrespective of their strength. However, in the
presence of a magnetic field new energy and length scales are introduced,
and they lead to nontrivial spectra. 

For regularly distributed impurities the problem can be handled by
introducing periodic solutions. For simplicity the square lattice case will
be investigated. Generalization to an arbitrary lattice is straightforward. 
Now, the impurity positions are given by 
${\bf r}_{mn}=d(m,n)$ where $m$ and $n$ are integers and $d$ is the lattice 
constant. The matrix elements become 
\begin{equation}
<\!\!mn|{\tilde V}|m'n'\!\!>=
\frac{V_0}{\pi}\exp[-\pi\frac{\phi}{\phi_0}
(\frac{(m-m')^2+(n-n')^2}{2}+i(mn'-nm'))]
\end{equation}
where $\phi$ is the magnetic flux passing through a unit cell 
(or flux per impurity) and 
$\phi_0$ is the flux quantum (note that the ratio of the area of the 
system to the number of states is $\pi$). When $\phi/\phi_0$ is a 
rational number the system 
becomes periodic. Let $\phi/\phi_0=\alpha=p/q$ where $p$ and $q$ are 
relatively
prime integers then $<\!\!mn|{\tilde V}|m'n'\!\!>$ is left invariant under 
$2q$
translations along lattice axes. If $p$ is even, the matrix elements 
become periodic with period $q$ instead of
$2q$. In any case, as long as $\alpha$ is rational, the eigenvectors
$|n{\bf k}\!\!>$ have the property that
\begin{equation}
<\!\!{\bf R}{\bf \tau}|n{\bf k}\!\!>=<\!\!0{\bf \tau}|n{\bf k}\!\!>
e^{i{\bf k}.{\bf R}}
\end{equation}
where ${\bf R}$ fixes the $2q\times 2q$ (or $q\times q$) unit cell and 
${\bf \tau}$ denotes the position of the impurity in the cell. Here
$n$ is an integer (band index) and ${\bf k}$ is a two dimensional vector. 
Hence, the problem is reduced to determination of the coefficients 
$<\!\!0{\bf \tau}|n{\bf k}\!\!>$ which satisfy
\begin{equation}
\sum_{\bf \tau}{\tilde V}_{{\bf \tau}{\bf \tau}'}({\bf k})
<\!\!0{\bf \tau}'|n{\bf k}\!\!>=E_{n{\bf k}}<\!\!0{\bf \tau}|n{\bf k}\!\!>
\end{equation}
where
\begin{equation}
{\tilde V}_{{\bf \tau}{\bf \tau}'}({\bf k})=\sum_{\bf R}
<\!\!{\bf R}{\bf \tau}|{\tilde V}|0{\bf \tau}'\!\!>e^{i{\bf k}.{\bf R}}~~~. 
\end{equation}
Given the flux ratio $\alpha$, the energy eigenvalues $E_{n{\bf k}}$ 
can be found by diagonalizing $4q^2\times 4q^2$ matrix 
${\tilde V}_{{\bf \tau}{\bf \tau}'}({\bf k})$. However, this does not imply 
that there will be $4q^2$ energy bands. If the problem is formulated in 
the Landau gauge, the additional symmetry can be seen easily. It turns out
that unit cell has dimensions $1\times q$ and hence  there will be $q$ bands. 
The symmetric gauge leads to a simple formulation of the problem but in 
the mean time some symmetries are obscured. 

In Fig.~\ref{band} energy bands for $\alpha=4/3$ and 
$\alpha=10/7$ are plotted. There are 3 and 7 energy bands, 
respectively. 
Using the relation $|\psi\!\!>=E^{-1/2}A|{\tilde \psi}\!\!>$, 
it is possible
to find the corresponding eigenfunction $\psi ({\bf r})$ of the original 
problem. Since $\psi ({\bf r})$ belongs to the lowest Landau level
it must be in the form of $f(z)e^{-|z|^2/2}$ where $f(z)$ is any analytic
function. Analytic functions are determined by their zeros. Considering the
translational invariance of the system one expects to have regularly
distributed zeros. Figure~\ref{zero} shows the contour plots of 
$|\psi ({\bf r})|$ for $\alpha=2$. This value of the flux ratio is very
special in that all of the matrix elements $<\!\!mn|{\tilde V}|m'n'\!\!>$ 
become
real and the problem reduces to a tight-binding system. For this particular
case, energy eigenvalues $E_{\bf k}$ and eigenfunctions $\psi_{\bf k}$ can
be found explicitly in terms of the theta function
\begin{equation}
\theta_3(z|\tau)=\sum_{n=-\infty}^{\infty}\exp(\pi i \tau n^2)
\exp(2n\pi iz)
\end{equation}
which vanishes at $z=(l+1/2)+(m+1/2)\tau$ where $l$ and $m$ are any
integers. It can be shown that
\begin{equation}
E_{\bf k}=\frac{V_0}{\pi}\theta_3(\frac{k_x}{\sqrt{2\pi}}|2i)
\theta_3(\frac{k_y}{\sqrt{2\pi}}|2i)
\end{equation}
and
\begin{equation}
\psi_{\bf k}(z, z^*)=ce^{-|z|^2/2}
\theta_3(\frac{z+ik_x}{i\sqrt{2\pi}}|i)
\theta_3(\frac{z-k_y}{\sqrt{2\pi}}|i)
\end{equation}
where $d$ is the lattice constant and $c$ is chosen to normalize 
$\psi_{\bf k}(z, z^*)$ properly. $\psi_{\bf k}(z, z^*)$ has regularly
distributed zeros. For ${\bf k}=0$, the zeros of the two $\theta_3$
functions coincide and they become double zeros. For ${\bf k}\neq 0$, they
shift by an amount $-ik_x$ and $k_y$, respectively.

Figure~\ref{periodic} shows energy (measured in units of $V_0/\pi$) versus 
the flux ratio for the square 
lattice. Here the labels $n$ and ${\bf k}$ have been suppressed, i.e,
$E_{n{\bf k}}$ values have been projected on to the energy axis. The energy
spectrum is found to be self-similar, i.e. the same pattern is repeated with
a different size and in a slightly distorted shape. With increasing flux
ratio, the spectrum is squeezed since the amplitude of the matrix elements
decreases with distance in a Gaussian way. 

In conclusion, a new method for the evaluation of the energy spectrum of 
a particle in the lowest Landau level moving in an arbitrary potential, 
is developed. The problem of Dirac delta scattering centers is solved 
for demonstration. The periodic case is found to have a self-similar
spectrum. The method is very general since it can be applied not 
only to any distribution of impurities but any potential, as far as the 
particle is confined into the lowest Landau level. For the random
distribution of scatterers\cite{gedik}, the calculated density of states 
is in perfect agreement with the analytic result of Brezin {\it et
al.}\cite{brezin}.
For a continuous 
potential, the matrix elements are labeled by continuous indices and hence 
the proposed approach may not be as practical as it is in the Dirac delta 
type potential case.

This work was partially supported by the Scientific and Technical Research
Council of Turkey under Grant No. TBAG-AY/123. One of the authors (Z.G.) is 
grateful to O. Tekman for helpful conversations.

\begin{figure}
\caption{Energy bands over one-quarter of the magnetic Brillouin zone for 
(a) $\alpha=4/3$ and (b) $\alpha=10/7$. The energy 
eigenvalues $E_{\bf k}$ are plotted in units of $V_0/\pi$.}
\label{band}
\end{figure}

\begin{figure}
\caption{Contour plots of $|\psi_{\bf k}({\bf r})|$ for $\alpha=2$. Zeros
are marked with *. (a) $k_x=k_y=0$, (b) $k_x=\sqrt{\pi/2}$ and $k_y=0$,
(c) $k_x=0$ and $k_y=\sqrt{\pi/2}$, (d) $k_x=k_y=\sqrt{\pi/2}$. }
\label{zero}
\end{figure}

\begin{figure}
\caption{Energy eigenvalues versus magnetic flux (in units of flux 
quantum) per unit cell for square lattice.}
\label{periodic}
\end{figure}

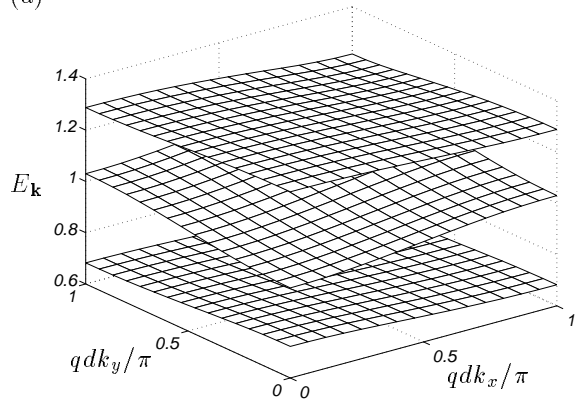
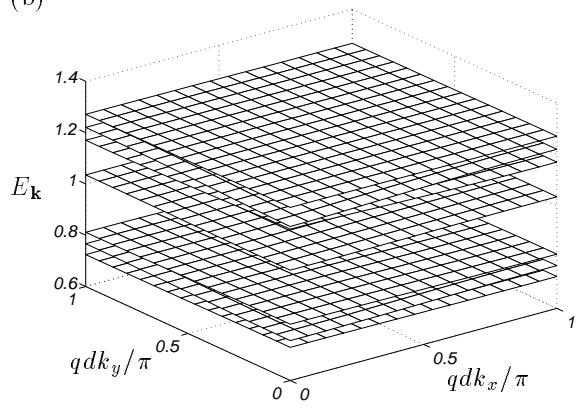
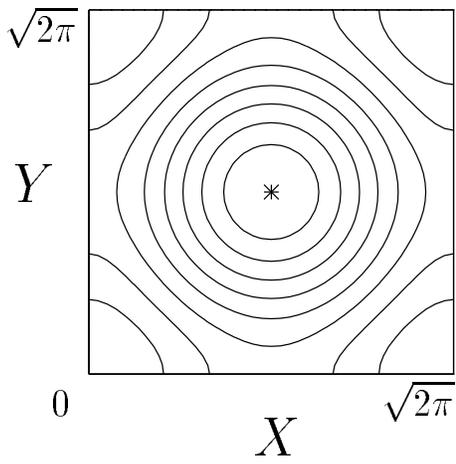
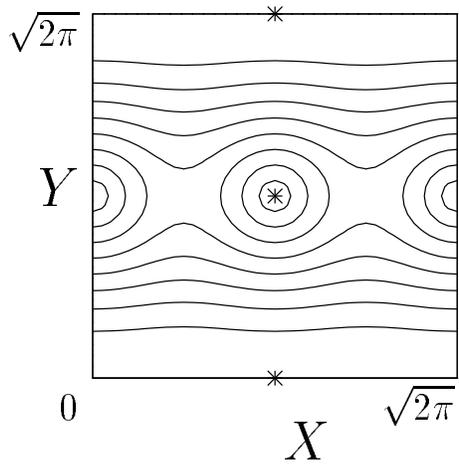
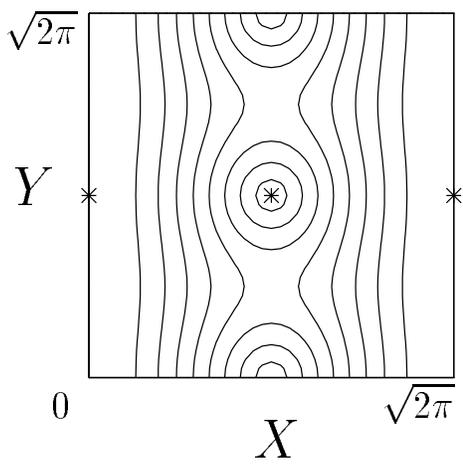
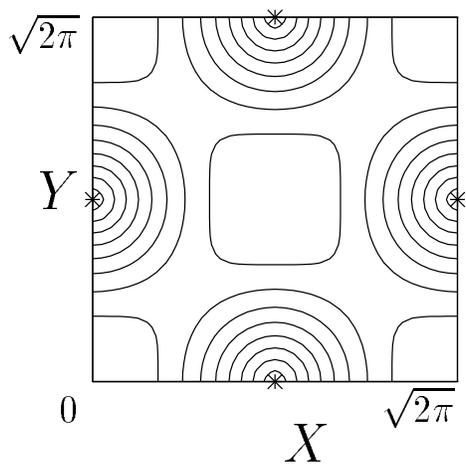
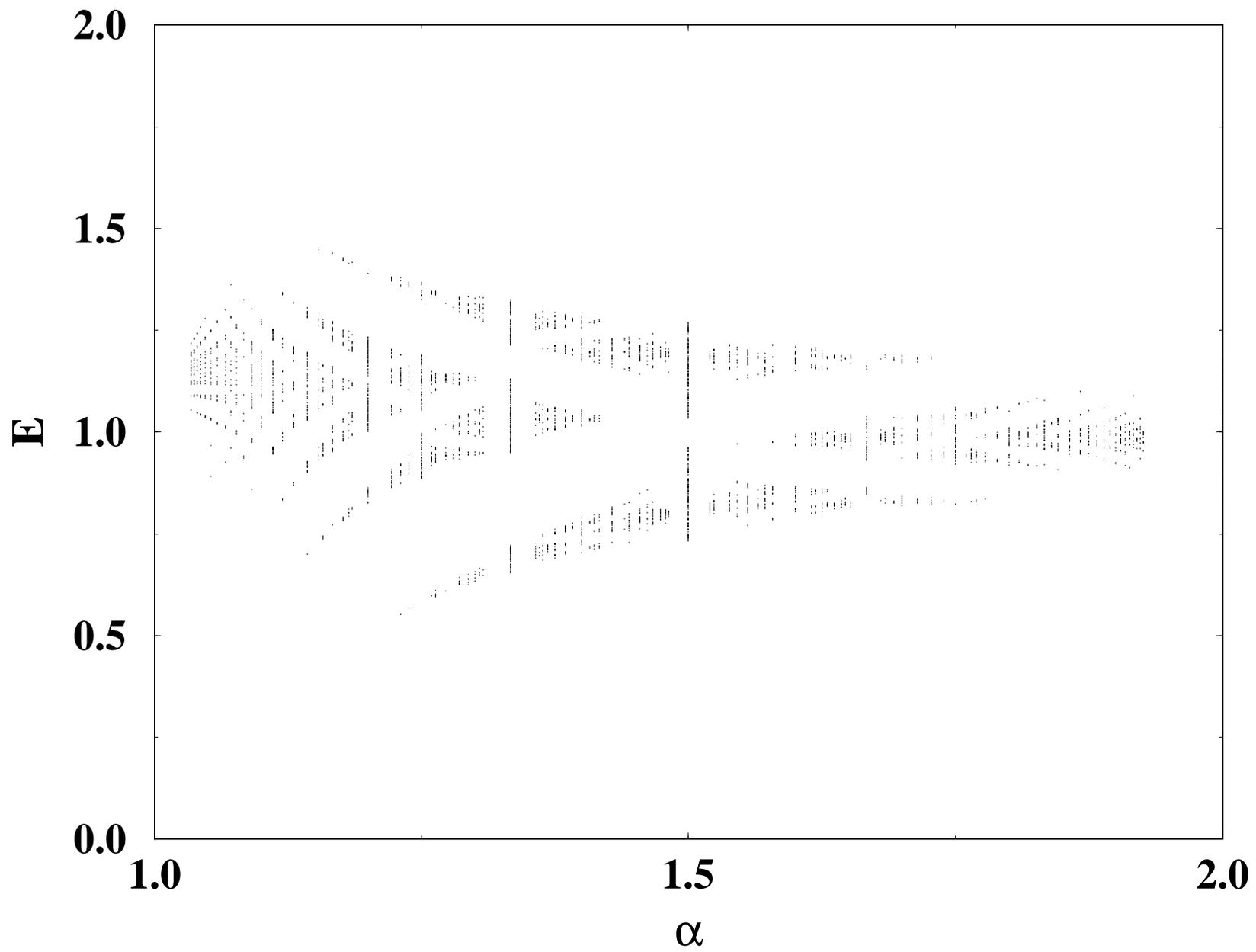
\end{document}